\documentclass{mem}
\usepackage{natbib}
\usepackage{txfonts}
\usepackage{balance}
\usepackage{graphicx}
\usepackage[a4paper,breaklinks,dvipdfm]{hyperref}
\idline{75}{282}

\begin{document}
\def\teff{$T\rm_{eff }$}
\def\kms{$\mathrm {km s}^{-1}$}

\title{
The substellar content of the \\
Orion Nebula Cluster
}

\author{H.\,Drass \inst{1} \and
          R.\,Chini\inst{1,2} \and
          D.\,N\"urnberger\inst{1} \and
          A.\,Bayo\inst{3,4} \and
          M.\,Haas\inst{1} \and
          M.\,Hackstein\inst{1} \and
          M.\,Morales-Calder\'on\inst{5} \and
          V.\,Hoffmeister\inst{1}   
          }
  \institute{
            Astronomisches Institut, Ruhr-Universit\"at Bochum, Germany \label{inst1} 
            \and
            Instituto de Astronom\'ia, Universidad Cat\'olica del Norte, Antofagasta, Chile\label{inst2}           
            \and
            European Southern Observatory, Santiago, Chile\label{inst3}
            \and
            Max-Planck Institut f\"ur Astronomie, Heidelberg, Germany  \label{inst4}         
            \and         
            Centro de Astrobiolog\'ia (INTA-CSIC), Departamento de Astrof\'isica, Madrid, Spain \label{inst7}
            }
            
  \offprints{hdrass@astro.rub.de}

  \authorrunning{H. Drass}

  \titlerunning{The substellar content of the ONC}

   \abstract
   {The Substellar Initial Mass Function (SIMF) of many star-forming regions is still poorly known but  the detailed knowledge of its shape will help to distinguish among the substellar formation theories. The Orion Nebula Cluster (ONC) is one of the most extensively studied star forming regions. We here present deep, wide-field $JHK$ observations of the ONC taken with HAWK-I@VLT. These observations extend the IMF into the brown dwarf and free-floating planetary mass regime with unprecedented sensitivity.
   To obtain a clean sample of ONC members, we exclude potential background sources with the help of CO extinction maps. Masses are assigned by means of evolutionary tracks in the \mbox{$H$ vs. $J-H$ Color-Magnitude Diagram (CMD)}. Besides the well known stellar peak at $\sim$0.25\,$M_{\odot}$ we find a pronounced second peak at $\sim$0.04\,$M_{\odot}$ in the SIMF and indications for a third rise in the free-floating planetary mass regime.
   }

\maketitle{}

\section{Introduction}
  Stars form over a wide range of masses with a factor of 1000 between stars of highest and lowest mass.
  Considering only the Jeans mass limit of a typical molecular cloud, substellar objects cannot form following the classical fragmentation/accretion scenario. To solve this situation, modified theories have been proposed. All these mechanisms would leave an imprint in the resulting IMF. Therefore, studying the low-mass end of the IMF is a valuable tool to investigate the formation of substellar objects.
  The nearby ($414\pm7$\,pc,  \mbox{Menten, 2007, A\&A, 474, 515}) Orion Nebular Cluster (ONC) with an age of only $\sim$1--\,3\,Myr (Hillenbrand, 1997, AJ, 113, 1733) is ideal to search for Brown Dwarf Candidates (BDCs) and Isolated Planetary Mass Object Candidates (IPMOCs).

\section{Membership}
  \label{mem_det}
  Deep surveys like our $JHK$--\,HAWK-I investigation run into confusion problems. Apart from the unavoidable presence of foreground and background stars, the position of Orion above the galactic disk (l\,$\sim$\,19$\,\deg$) has got the disadvantage that a priori nothing shields extragalactic sources from contributing light at NIR wavelengths except Orion's own molecular cloud.
  The Orion Nebula and the associated ONC are located at the front-side of the optically thick \mbox{OMC-1} molecular cloud. Therefore, comparing the visual extinction for an individual source with the total extinction of the cloud along this particular line of sight should allow a discrimination of background sources. Foreground contribution from nearby cluster offshoots are not excluded by this approach.

  Scandariato et al. (2011, A\&A, 533, A38) used a large number of bona fide background stars to obtain an extinction map of the \mbox{OMC-1} with an angular resolution $< 5\arcmin$.
  \mbox{Shimajiri et al. (2011, PASJ 63, 105)} mapped the region in $^{12}$CO with a spatial resolution of 21\arcsec. Neglecting any effects of optical depth the CO map can be calibrated with the optical extinction derived by \mbox{Scandariato et al. (2011, A\&A, 533, A38)}. Overall, the relation shows an linear dependency of $A_V$ with CO intensity.

  Next we determine $A_V$ for 4107 sources from our master catalog by investigating their location in the $H$ vs. $(J-H)$ diagram.

\begin{figure}
      \includegraphics[width=\hsize, keepaspectratio, angle=0, trim=0cm 0cm 0cm 0cm, clip=false]{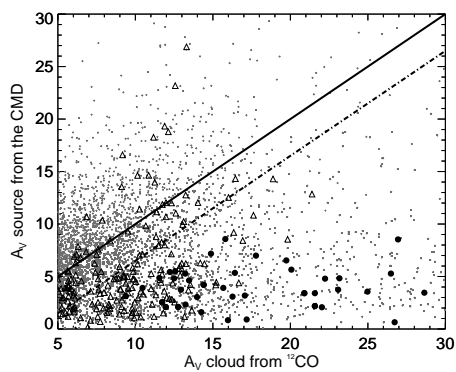}
      \caption{$A_V$ derived from the $(J-H)$ color excess vs. the $A_V$ from our CO calibration. The offset between the solid and dash-dotted lines corresponds to $A_V = 3.5$\,mag. Open triangles denote the objects with YSOVAR counterparts. Known Brown Dwarfs are marked with filled circles.
      }
      \label{AV_src_vs_AV_CO_TIO}
  \end{figure}

  In Fig.~\ref{AV_src_vs_AV_CO_TIO} we have plotted the resulting $A_V$ values of our stars versus the corresponding $A_V$ value as obtained from the CO data. We consider all stars below the solid line as member candidates (large sample) because their $A_V$ is lower than expected for background stars along the particular line of sight. The upper envelope of this population is shown as a dashed-dotted line highlighting 3.5\,mag of extinction lower then the limit marked by the solid line (conservative sample).

  To check the reliability of our procedure we compare the behavior of bona fide YSOs to the sources reported in by the YSOVAR survey (Morales-Calder\'on 2011,  ApJ 733, 50).
  $82\%$ YSOVAR sources display extinction values in agreement with our membership criterion while $18\%$ disagree.

  A striking test can be made by means of confirmed BDs in the ONC (Slesnick 2004, ApJ, 610, 1045; Riddick 2007, MNRAS, 381, 1077; Weights 2009, MNRAS, 392, 817). Every confirmed BD lie in the "member area" of Fig.~\ref{AV_src_vs_AV_CO_TIO} providing further confidence that our extinction-based membership criterion yields reliable results.

\section{Mass determination}
\begin{figure}
    \resizebox{\hsize}{!}{\includegraphics[ trim=1.5cm 0cm 0cm 0cm, clip=true]{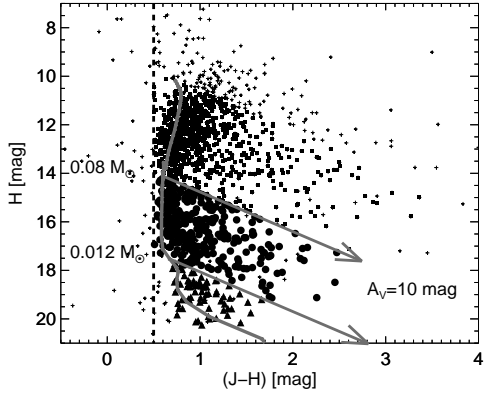}}
    \caption{Color-magnitude diagram for cluster members from the conservative sample.
    Dashed line at $(J-H)_0 = 0.5$ corresponds to the intrinsic color of late-type main sequence stars; the curve denotes the 3\,Myr isochrone up to $1.4\,M_{\odot}$.}
    \label{CMD_H_vs_JmH}
  \end{figure}

  Fig.~\ref{CMD_H_vs_JmH} displays the color-magnitude diagram for those stars that classified as cluster members.
  The result is consistent with previous results (Hillenbrand \& Carpenter, 2000, ApJ, 540, 236; Da Rio 2010, ApJ, 722, 1092) but includes $\sim$1.5\,mag fainter objects.
  The selection contains only sources detected in $J,H$ and $K$ with a $S/N >3$ and a PSF photometric error $<0.1$\,mag.

  We derive mass estimates by dereddening the objects and shifting them back to the 3\,Myr isochrone from Allard et al. (2011, ASPC 448, 91) (Fig.~\ref{CMD_H_vs_JmH}).
  Masses could be determined for $92\%$ of both samples.
  Sources from the conservative sample are denoted by different symbols: $1130$ low-mass stars (squares), $300$ BDs (dots) and $70$ IPMO candidates (triangles).
  For the large sample $656$ sources fall in the BD range while $352$ sources are IPMO candidates; the remaining sources have stellar masses.

\section{Initial Mass Function}
\begin{figure}
      \centering
      \resizebox{\hsize}{!}{\includegraphics[angle=0, trim=1.2cm 0cm 0cm 0cm, clip=false]{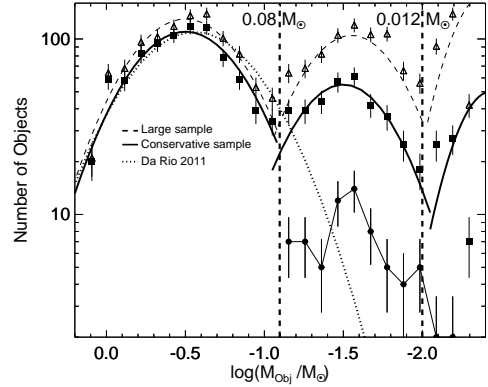}}
      \caption{Initial mass function for members in the range of Allard's 3\,Myr isochrone. Solid curves are fitted to the conservative sample, the dashed curves show the large sample; BDs are located within the two dash-dotted lines. For comparison the dashed-dotted curve displays the result by Da Rio (2012, ApJ, 748, 14).
      Filled circles are the known BDs. Uncertainties are given by $\sqrt {N}$, where $N$ is the number of sources per bin.
      }
      \label{IMF_excess}
    \end{figure}

  With the described mass estimates we compile the mass distribution by binning the member candidates.
  The large number of objects allows us to have equally-sized mass bins with a statistically significant amount of objects in each bin to minimize binning effects on our conclusions.
  The resulting IMF (Fig.~\ref{IMF_excess}) shows two statistically significant bumps and a final increase towards faint sources. All features become more prominent for the large sample. There are two pronounced dips at the BD ($0.08\,M_\odot$) and the IPMO ($0.01\,M_\odot$) limit whose origin is unclear.
  For the masses below 0.01$M_{\odot}$ we would like to note that uncertainties in the mass determination are dominated by the fact that the isochrone is nearly parallel to the reddening vector in the CMD. As a consequence the IMF rise in the IPMOCs range is more of indicative character.
  The uncertainty can be estimated from the YSOVAR objects in Fig.~\ref{AV_src_vs_AV_CO_TIO} that yields a contamination of \mbox{$\sim20\%$} for this mass range.
  Nevertheless this is not changing our results significantly, especially considering the large number of sources in the large sample.
  To further support our results we plot the mass function of the few known BDs (Fig.~\ref{IMF_excess}) and also find indications for a peak at the same mass than our sources. We also plot the results derived by Da Rio et al. (2012) showing a steep decline in the BD mass range.

\section{Conclusions}
  We investigated the substellar population of the ONC.
  As a novel approach to provide new candidates we proposed to disentangle members from background objects employing an extinction map of the \mbox{OMC-1} derived from molecular and stellar data and compare this with the extinction for each source. We found a suitable relation to conservatively select ONC member candidates.
  We conclude, that a unique classification of both the nature and membership of the substellar candidates requires spectroscopic information.
  
  \begin{acknowledgements}
  H. Drass was funded by ESO, Chile during a two year studentship.
  A. Bayo was partially funded under the Marie Curie Actions of the European Commission (FP7-COFUND).
\end{acknowledgements}

\end{document}